\documentclass[prl,twocolumn,amsmath,amssymb,superscriptaddress, bibnotes]{revtex4-2}

\usepackage{graphicx}
\usepackage{dcolumn}
\usepackage{bm}
\usepackage{color}

\begin{document}

\title{Parity transition of spin-singlet superconductivity using sub-lattice degrees of freedom}

\author{Shiki~Ogata}

\affiliation{Department of Physics, Kyoto University, Kyoto 606-8502, Japan}

\author{Shunsaku~Kitagawa}

\email{kitagawa.shunsaku.8u@kyoto-u.ac.jp}

\author{Katsuki~Kinjo}
\author{Kenji~Ishida}

\affiliation{Department of Physics, Kyoto University, Kyoto 606-8502, Japan}

\author{Manuel~Brando}
\affiliation{Max Planck Institute for Chemical Physics of Solids, D-01187 Dresden, Germany}

\author{Elena~Hassinger}
\affiliation{Technical University Dresden, Institute for Solid State and Materials Physics, 01062 Dresden, Germany}

\author{Christoph~Geibel}
\author{Seunghyun~Khim}

\affiliation{Max Planck Institute for Chemical Physics of Solids, D-01187 Dresden, Germany}

\date{\today}

\begin{abstract}
Recently, a superconducting (SC) transition from low-field (LF) to high-field (HF) SC states was reported in CeRh$_2$As$_2$, indicating the existence of multiple SC states.
It has been theoretically noted that the existence of two Ce sites in the unit cell, the so-called sub-lattice degrees of freedom owing to the local inversion symmetry breaking at the Ce sites, can lead to the appearance of multiple SC phases even under an interaction inducing spin-singlet superconductivity.
CeRh$_2$As$_2$ is considered as the first example of multiple SC phases owing to this sub-lattice degree of freedom.
However, microscopic information about the SC states has not yet been reported.
In this study, we measured the SC spin susceptibility at two crystallographically inequivalent As sites using nuclear magnetic resonance for various magnetic fields.
Our experimental results strongly indicate a spin-singlet state in both SC phases.
In addition, the antiferromagnetic phase, which appears within the SC phase, only coexists with the LF SC phase; there is no sign of magnetic ordering in the HF SC phase.
The present work reveals unique SC properties originating from the locally noncentrosymmetric characteristics.
\end{abstract}

\maketitle

\begin{figure*}[!tb]
\includegraphics[width=14cm,clip]{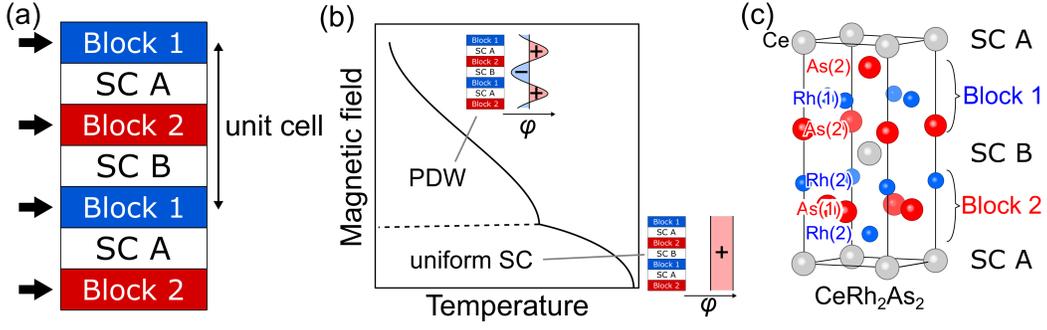}
\caption{
Schematics of a pair-density-wave (PDW) state in a two-dimensional system with locally broken inversion symmetry.
(a) Schematic image of crystal structure.
The unit cell contains two kinds of superconducting layers.
The position of the inversion center is indicated by the arrows.
(b) Theoretically predicted $H$ --$T$ phase diagram.
The position dependence of superconducting order parameters in each superconducting phase is shown by the schematic images.
(c) Crystal structure of CeRh$_2$As$_2$. 
}
\label{Fig.1}
\end{figure*}

In conventional superconductors, a superconducting (SC) order parameter can be classified as the multiplication of spin and orbital states, that is, even-parity spin-singlet and odd-parity spin-triplet states. 
One possible expansion of the framework in SC symmetry is to introduce additional degrees of freedom, such as frequency and atomic orbital\cite{J.Linder_RMP_2019,H.G.Suh_PRR_2020}.
As shown in Fig.~\ref{Fig.1}(a), in a two-dimensional system with locally broken inversion symmetry (centrosymmetric but no inversion center at the SC layers), there are sub-lattice degrees of freedom owing to the inversion symmetry operation between the two layers. 
In such systems, a parity transition from a low-field (LF) even-parity SC state to a high-field (HF) odd-parity pair-density-wave (PDW) SC state (the SC phase is inverted layer by layer) [Fig.~\ref{Fig.1}(b)] is theoretically proposed, even when only a pairing interaction in the spin-singlet channel exists\cite{T.Yoshida_PRB_2012}.
The PDW state coexisting with the charge-density-wave state has been discussed in high-SC transition temperature $T_{\rm SC}$ cuprates\cite{D.F.Agterberg_ARCMP_2020} and kagom\'{e} superconductors\cite{H.Chen_Nature_2021}; however, the PDW phase originating from crystal symmetry is quite rare.

CeRh$_2$As$_2$, where the local inversion symmetry is broken at the Ce sites, is a recently discovered superconductor with $T_{\rm SC} \sim 0.3$~K\cite{S.Khim_Science_2021}.
In CeRh$_2$As$_2$, an SC transition from the SC1 to SC2 phase occurring at approximately 4~T was reported for $H \parallel c$.
This is the first experimental indication of the SC parity transition originating from the sub-lattice degrees of freedom and has promoted many theoretical studies\cite{E.G.Schertenleib_PRR_2021,K.Nogaki_PRR_2021,A.Skurativska_PRR_2021,D.Mockli_PRR_2021,D.Mockli_PRB_2021,A.Ptok_PRB_2021,D.C.Cavanagh_PRB_2022}.
However, to date, there have been few experimental investigations on the SC properties\cite{M.Kibune_PRL_2022,S.Onishi_FEM_2022,J.F.Landaeta_arXiv_2022}.
Therefore, further experimental confirmation is required.

The crystal structure of CeRh$_2$As$_2$ is of the tetragonal CaBe$_{2}$Ge$_{2}$-type with space group $P4/nmm$ (No.129, $D_{4h}^7$)\cite{R.Madar_JLCM_1987}.
There are two crystallographically inequivalent As and Rh sites.
As(1) [Rh(1)] is tetrahedrally coordinated by Rh(2) [As(2)], as shown in Fig.~\ref{Fig.1} (c).
Heavy-fermion superconductivity is characterized by a broad maximum in resistivity at $T_{\rm coh} \sim$ 40~K and a large specific heat jump at $T_{\rm SC}$\cite{S.Khim_Science_2021}.
In addition to the multiple SC phases, CeRh$_2$As$_2$ exhibits nonmagnetic and magnetic phase transitions just above and below $T_{\rm SC}$, respectively\cite{S.Khim_Science_2021,M.Kibune_PRL_2022}.
The specific heat shows a large anomaly at $T_{\rm SC}$ and a rather weak anomaly at $T_0 \sim$ 0.4~K.
The anomaly at $T_0$ is considered as a phase transition to an electric quadrupole density-wave state\cite{D.Hafner_PRX_2022}.
Moreover, nuclear quadrupole resonance (NQR) measurements revealed an antiferromagnetic (AFM) order inside the SC phase\cite{M.Kibune_PRL_2022}.
Hence, CeRh$_{2}$As$_{2}$ is a promising system for studying the role of sub-lattice degrees of freedom for unconventional nonmagnetic, AFM, and SC states as well as their interactions.

\begin{figure*}[!tb]
\includegraphics[width=18cm,clip]{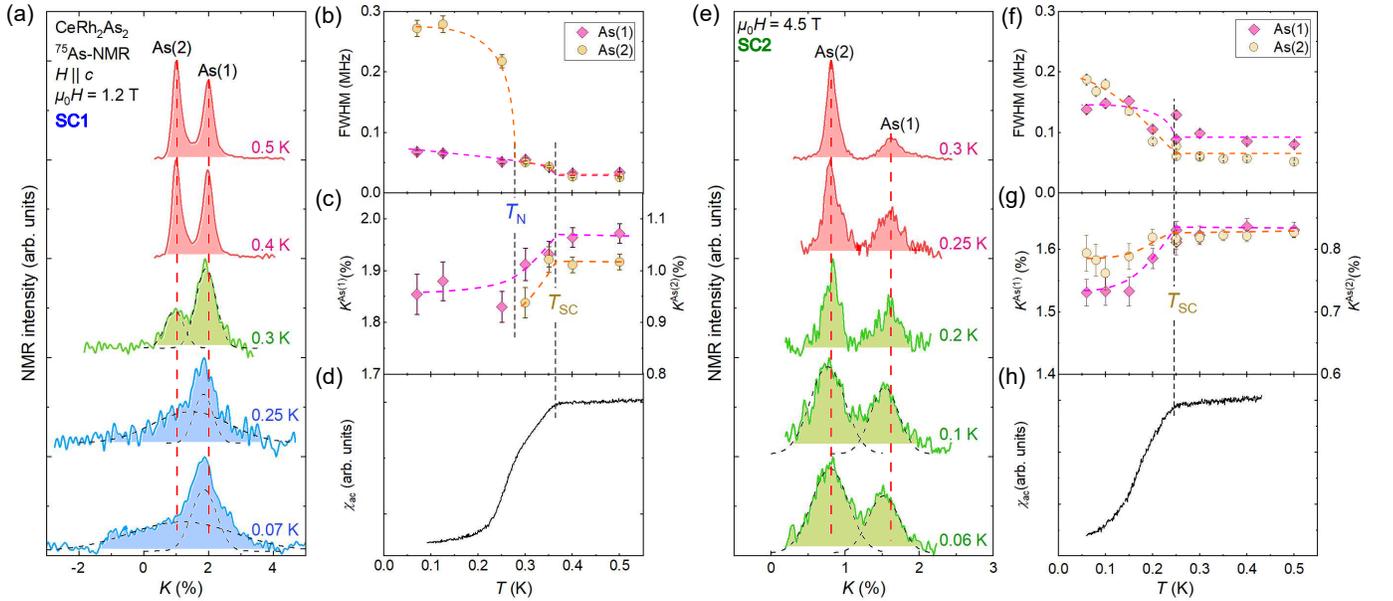}
\caption{
NMR measurements in CeRh$_2$As$_2$.
(a) NMR spectra at 1.2~T for $H \parallel c$ measured at various temperatures.
The dashed curves indicate the results of a two-peak fitting.
Temperature dependence of full width at half maximum(FWHM) (b), Knight shift (c) at the As(1) site, $K^{\rm As(1)}$, and As(2) site, $K^{\rm As(2)}$, determined by NMR spectrum at 1.2~T.
$T_{\rm SC}$ and $T_{\rm N}$ are indicated by the dashed lines.
The dashed curves are guides for the eye.
(d) Temperature dependence of ac magnetic susceptibility at 1.2~T for $H \parallel c$.
(e) NMR spectra at 4.5~T for $H \parallel c$ measured at various temperatures.
Temperature dependence of FWHM (f), Knight shift (g), and ac magnetic susceptibility (h) at 4.5~T for $H \parallel c$.
}
\label{Fig.2}
\end{figure*}

Nuclear magnetic resonance (NMR) can measure spin susceptibility in the SC state, whereas bulk magnetic susceptibility is dominated by SC diamagnetic shielding effect.
In addition, NMR is sensitive to the appearance of internal magnetic fields at observed nuclear sites.
Therefore, NMR is one of the most powerful techniques for studying SC and magnetic properties.
Furthermore, NMR is also useful for detecting spatially modulated SC states, such as the Fulde-Ferrell-Larkin-Ovchinnikov and PDW phases, because NMR measures the local spin susceptibility at the nuclear positions\cite{K.Kinjo_Science_2022}.

In this paper, we report the $^{75}$As-NMR results for CeRh$_2$As$_2$.
Up to 5~T, the spin susceptibility decreased in the SC state, indicating spin-singlet superconductivity in both the SC1 and SC2 phases.
The decrease in the Knight shift in the SC1 phase was in good agreement with the critical field of the SC1 -- SC2 transition.
In contrast, the decrease in the Knight shift in the SC2 phase is seemingly inconsistent with the large $H_{\rm c2}$, suggesting the spatial modulation of the SC spin susceptibility.
This is the first microscopic information on the unique SC states in CeRh$_{2}$As$_{2}$.

Single crystals of CeRh$_{2}$As$_{2}$ were grown using the Bi flux method\cite{S.Khim_Science_2021}.
The magnetic field dependence of $T_{\rm SC}$ and the temperature dependence of the critical field of the SC1 -- SC2 transition were determined from the ac susceptibility measurements using an NMR coil as shown in Supplemental Materials (SM)\cite{SM}.
The details in NMR measurements are described in SM\cite{SM}.
We experimentally confirmed the superconductivity immediately after the NMR pulses using a technique reported in a previous study\cite{K.Ishida_JPSJ_2020,H.Fujibayashi_JPSJ_2022}.
Reflecting two crystallographically inequivalent As sites, two NMR peaks were observed in all measurement ranges, as shown in Figs.~\ref{Fig.2} (a) and \ref{Fig.2} (e).
The site assignment of two NMR peaks has been described in a previous paper\cite{S.Kitagawa_JPSJ_2022}.
The reason why the NMR spectrum of the As(1) site is broad at 4.5 T is a slight misalignment of the magnetic field direction from the $c$-axis ($<$2$^{o}$ in this experiment).

\begin{figure*}[!tb]
\includegraphics[width=14cm,clip]{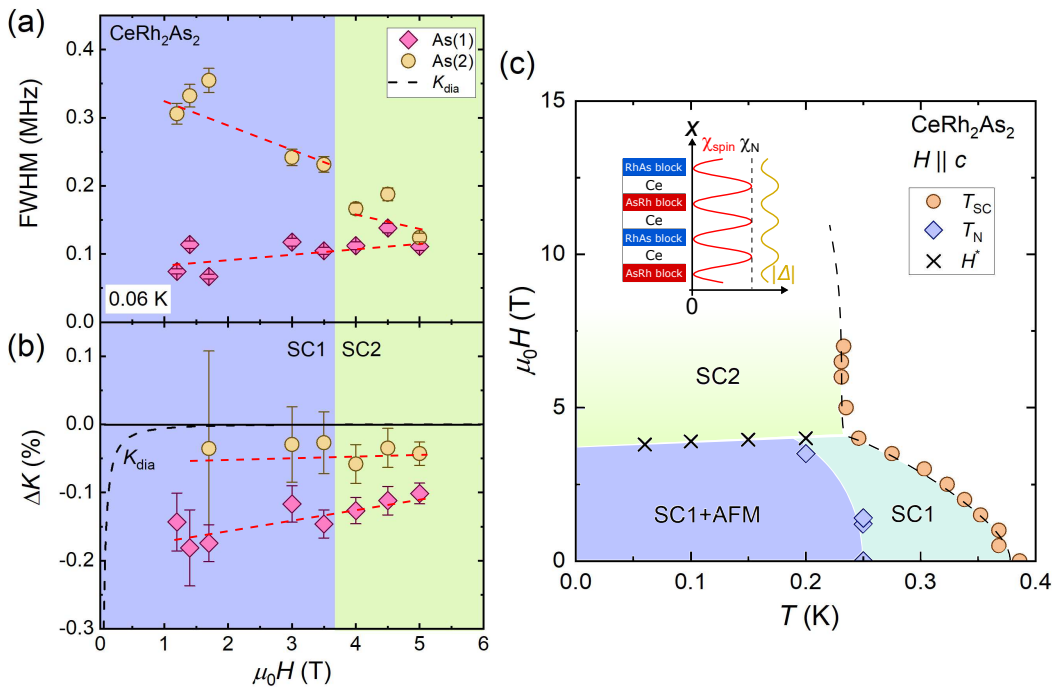}
\caption{
Magnetic field variation of the magnetic properties in CeRh$_2$As$_2$.
The $c$-axis magnetic field dependence of full width at half maximum (FWHM) at 0.06~K (a) and $\Delta K~=$ $K$(0.06~K) - $K$(0.6~K) (b).
The dashed curve indicates the contribution of SC diamagnetism $K_{\rm dia}$.
The dashed lines are guides for the eye.
(c) The $H$ -- $T$ phase diagram in the present sample of CeRh$_2$As$_2$ determined from the NMR and ac magnetic susceptibility measurements.
Circles denote $T_{\rm SC}$ determined by the onset temperature of the SC diamagnetic signal from ac magnetic susceptibility.
Crosses denote the critical magnetic field of the SC1 -- SC2 transition determined by the anomaly of ac magnetic susceptibility.
Diamonds indicate $T_{\rm N}$ determined by the increase in NMR linewidth.
(Inset) Schematic image of the distribution of spin susceptibility and SC gap.
}
\label{Fig.3}
\end{figure*}

First, we show the temperature evolution of the NMR spectrum for each SC phase.
As shown in Fig~\ref{Fig.2} (a), at 1.2~T (low-field SC1 phase), two sharp peaks were observed above $T_{\rm SC} \sim 0.35$~K.
Below $T_{\rm SC}$, the NMR spectrum broadens, owing to the SC diamagnetic field, and significant site-dependent broadening is observed below 0.25~K.
The site-dependent broadening was also observed at zero field (NQR measurements)\cite{M.Kibune_PRL_2022}, which was attributed to the AFM order.
The internal field at the As(2) site is estimated to be $\sim$ 20~mT.
In contrast, at 4.5~T in the SC2 phase, the NMR lines broaden just below $T_{\rm SC}$ for both sites, as in the SC1 phase, but there is no additional, site-dependent broadening at lower temperatures, down to 0.06~K.
This indicates the absence of AFM ordering below $T_{\rm SC}$ in the SC2 phase.
To investigate the spin susceptibility at the two As sites, NMR Knight shifts are determined by two-peak Gaussian fitting.
The Knight shifts for the As(1) and As(2) sites are denoted by $K^{\rm As(1)}$ and $K^{\rm As(2)}$, respectively.
$K^{\rm As(1)}$ decreases below $T_{\rm SC}$ in both SC phases.
In the SC2 phase, $K^{\rm As(2)}$ also decreases, while in the SC1 phase $K^{\rm As(2)}$ is not measurable owing to the extremely broad spectrum below the AFM transition temperature $T_{\rm N}$.

To clarify the magnetic field evolution of the spin susceptibility and the AFM order, the magnetic field dependence of the full width at half maximum (FWHM) at the lowest temperature (0.06~K) and $\Delta K~=$ $K$(0.06~K)-$K$(0.6~K) are shown in Figs.~\ref{Fig.3} (a) and \ref{Fig.3} (b), respectively.
In the SC1 phase, the linewidth of the As(2) site decreases as the magnetic field increases but remains significantly broader than that of As(1) until 3.5~T.
Between 3.5~T and 4~T, the width of the As(2) lines drops to values only slightly higher than that of As(1).
In contrast, in the SC2 phase, the linewidth of the As(2) site is almost the same as that of the As(1) site.
The absence of an internal field at the As(2) site indicates that the AFM phase only coexists with the SC1 phase, not with the SC2 phase.
In contrast to the magnetic-field dependence of the linewidth, $|\Delta K|$ at both As sites does not show any anomaly at the phase boundary and gradually decreases as the magnetic field increases, up to 5~T.
The magnetic field dependence of $|\Delta K^{\rm As(1)}|$ can be extrapolated to zero at 14~T ($H_{\rm c2}$), suggesting that the magnetic field dependence of $|\Delta K|$ originates from the field-induced quasiparticle density of states (Volovik effect).
$|\Delta K^{\rm As(2)}|$ is smaller than $|\Delta K^{\rm As(1)}|$ owing to the difference of the hyperfine coupling constant $A_{\rm hf}$\cite{S.Kitagawa_JPSJ_2022}.

From the AC magnetic susceptibility and NMR measurements, we constructed the $H$ -- $T$ phase diagram in the present sample of CeRh$_2$As$_2$, as shown in Fig.~\ref{Fig.3} (c).
The previously reported peculiar shape of $T_{\rm SC}$ and the SC1 -- SC2 transition were reproduced by our measurements, whereas $T_{\rm SC} = 0.37$~K was higher than that of the previous samples\cite{S.Khim_Science_2021,J.F.Landaeta_arXiv_2022}.
This is due to the differences in sample quality and experimental methods.
Our NMR results indicate that the AFM phase disappears in the SC2 phase, which is strong evidence of the presence of the SC1 -- SC2 transition and the different SC properties between the two phases.
Furthermore, the smooth connection of the Knight shift between two SC states indicates that the SC2 state is a non-polarized one.
This is in complete contrast to the usual field-induced destruction of an AFM state resulting in a polarized ferromagnetic-like state.
Therefore, the observed field-induced transition from an AFM state to a state with no sizable polarization at high field is a highly unusual one, indicating that this AFM state is strongly linked to the SC1 state.

The SC spin state can be deduced from the temperature variation in the NMR Knight shift.
In a conventional spin-singlet superconductor, the spin component of the Knight shift decreases in the SC state and becomes almost zero at $T \rightarrow 0$~K.
However, when the SC spins of spin-triplet superconductors are aligned along the magnetic field, the spin component of the Knight shift does not change across $T_{\rm SC}$.
As the Knight shift includes various contributions, such as the temperature-independent orbital component arising from the Van Vleck susceptibility and SC diamagnetic effect, the absolute amount of the spin component of the Knight shift must be evaluated.
$K$ at the lowest temperature can be divided into the following three contributions:
\begin{align}
    K = K_{\rm normal} + \delta K_{\rm spin} + K_{\rm dia}.
\end{align}
where $K_{\rm normal}$ is the normal-state Knight shift, $\delta K_{\rm spin}$ is the change in a spin component of the Knight shift in the SC state, and $K_{\rm dia}$ is the SC diamagnetic component.
In the SC state, the Knight shift decreases owing to the SC diamagnetic shielding effect, and the value of $K_{\rm dia}$ at 0~K is approximately expressed as\cite{deGennes}
\begin{align}
    K_{\rm dia} = - \frac{H_{c1}}{H}\frac{\ln \left(\frac{\beta \lambda_d}{\sqrt{e}\xi}\right)}{\ln \kappa}.
\end{align}
Here, $H_{\rm c1}$ is the SC lower critical field; $\xi$ is the Ginzburg-Landau (GL) coherence length; $\beta$ is a factor that depends on the vortex structure and is 0.38 for the triangular vortex lattice; $\lambda_d$ is the distance between the vortices and is calculated using the relation $\phi_0 = \frac{\sqrt{3}}{2}\lambda_d^2 (\mu_0 H_{\rm ext})$; $e$ is Euler's number, and $\kappa$ is the GL parameter.
From the SC upper critical field $\mu_0 H_{\rm c2}~=~14$~T and thermodynamic critical field $\mu_0 H_{\rm c}~=~31$~mT\cite{S.Khim_Science_2021}, $\mu_0 H_{\rm c1}~=~0.40$~mT, $\xi~=~4.85$~nm, and $\kappa~=~319$ were obtained.
As shown in Fig~\ref{Fig.3} (b), the magnetic field dependence of $K_{\rm dia}$ is negligibly small compared to the experimental $\delta K$.
In addition, the temperature dependence of $K_{\rm normal}$ shows almost constant at low temperatures\cite{S.Kitagawa_JPSJ_2022}, as shown in Figs.~\ref{Fig.2} (c) and \ref{Fig.2} (g).
Therefore,  $\Delta K$ was dominated by $\delta K_{\rm spin}$.
As shown in Figs.~\ref{Fig.2} (c) and ~\ref{Fig.2} (g), a clear decrease in $K$ was observed in the SC state, indicating the realization of spin-singlet superconductivity in both the SC1 and SC2 phases.

The spin-singlet superconductivity in the SC1 phase can also be confirmed by the quantitative agreement between $H_{\rm c2}$ and the Pauli limiting field $H_{\rm P}$ estimated from the reduction in the Knight shift.
In strongly correlated electron systems, spin-singlet superconductivity is destroyed when the Zeeman-splitting energy is as high as the SC condensation energy, called the Pauli limiting effect.
It is well known that in a spin-singlet superconductor, a relation holds between $H_{\rm P}$ and the decrease in the spin susceptibility $\delta \chi$ ascribed to singlet-pair formation.
This is expressed as
\begin{align}
\frac{1}{2}\delta\chi \mu_0H_{\rm P}(0)^2~=~\frac{1}{2}\mu_0H_{\rm c}^2.
\end{align}
This equation yields $\mu_0H_{\rm P}(0) = \mu_0H_{\rm c}/\sqrt{|\delta \chi|}$, where $\delta \chi$ can be determined from $\Delta K$ as $\delta \chi = (\mu_{\rm B} N_{\rm A} / A_{\rm hf}) \Delta K$.
In the SC1 phase, $\Delta K$ at the As(1) site is approximately - 0.2 \%; thus, $\mu_0H_{\rm P}(0)$ is estimated to be 3.4~T with $\mu_0 H_{\rm c}$ = 31~mT and the hyperfine coupling constant $A_{\rm hf}^{\rm As(1)}$ =  1.55~T/$\mu_{\rm B}$\cite{S.Kitagawa_JPSJ_2022}. 
The estimated $\mu_0H_{\rm P}(0)$ agrees to the SC1--SC2 transition field ($\sim 4$~T), indicating that the SC1 phase exhibits homogeneous spin-singlet superconductivity.
Such a correspondence between the critical field and $\Delta K$ is observed in various spin-singlet heavy-fermion superconductors\cite{H.Tou_JPSJ_2005,S.Kitagawa_JPSJ_2017,S.Kitagawa_PRL_2018,T.Hattori_PRL_2018}.

In contrast, $\Delta K$ in the SC2 phase is unusual.
$K$ in the SC2 phase also clearly decreases, and $\mu_0H_{\rm P}(0)$ is estimated to be 4.8~T from $\Delta K^{\rm As(1)}~=~-0.1$ \%, which is much smaller than $\mu_0H_{\rm c2}(0) = 14$~T.
Here, we assume the same $H_{\rm c}$ as in the SC1 state.
If we consider that $H_{\rm P}(0)$ increases owing to an increase in $H_{\rm c}$ in the SC2 phase, the $H_{\rm c}$ should be increased by a factor of three, which is an unrealistic value in this case.
It is noteworthy that $\Delta K^{\rm As(2)}~=~-0.05$ \% and $A_{\rm hf}^{\rm As(2)}$ =  0.27~T/$\mu_{\rm B}$ also lead to a small Pauli-limiting field, $\mu_0H_{\rm P}(0)~=~2.8$~T.
There are several superconductors without the Pauli limiting effect owing to the absence of the spin susceptibility reduction\cite{H.Mukuda_PRL_2010,M.Manago_JPSJ_2017,G.Nakamine_JPSJ_2019}, but the case where the Pauli limiting effect does not work despite a decrease in the spin susceptibility is quite rare.
The discrepancy between the decrease in the spin susceptibility and the absence of the Pauli limiting field can be reconciled by the presence of a spatially inhomogeneous SC state.

Recently, SC states realized in the presence of sub-lattice degrees of freedom have been intensively studied, and spin susceptibility has been calculated in various SC states\cite{D.Maruyama_JPSJ_2013}.
The average spin susceptibility in the PDW state is suggested to be the same as that in the normal state.
The present Knight-shift results are inconsistent with these theoretical suggestions. 
To interpret this discrepancy, we propose a position-dependent modulation of spin susceptibility in the PDW state. 
In this SC state, the SC order parameter oscillates along the $c$ axis with a period corresponding to the $c$ lattice parameter; thus, the spin susceptibility also oscillates with the same periodicity. 
NMR Knight shift probes the local spin susceptibility at the observed nuclear site.
Therefore, the present results imply that the block layers become the "spin-singlet" dominant SC state and the spin susceptibility at the block layers decreases in the SC2 states as shown in the inset of Fig.~\ref{Fig.3} (c). 
In contrast, a sizable Rashba effect at the Ce site would lead to locally spin-polarized bands. 
The Pauli susceptibility would then transform into a Van Vleck susceptibility, which is insensitive to superconductivity\cite{D.Maruyama_JPSJ_2013,D.Maruyama_JPSJ_2012}.
As a result, the spin susceptibility in the SC state at the Ce sites would be of similar size as that in the normal state.

In CeRh$_2$As$_2$, the SC properties are determined by the Ce layer, resulting in the spin susceptibility of the Ce layers being crucially modified by the SC character. 
We suggest the possibility that the spin susceptibility at the Ce layers largely changes between the LF and HF SC states; that is, the spin susceptibility at the Ce site decreases in the SC1 state but remains unchanged in the SC2 state, although the spin susceptibility at the block layers does not change significantly. 
This might explain the large difference in the Pauli limiting field between the two phases. 
However, because all NMR-active Ce isotopes are unstable, measuring the spin susceptibility at the Ce site is impossible.
Thus, we need to seek other methods to determine the spatial modulation of the spin susceptibility in CeRh$_2$As$_2$.
In addition, theoretical studies on spin-susceptibility modulation in the PDW state are required.

In conclusion, we measured the SC spin susceptibility in the LF and HF SC states.
The spin susceptibility decreases in both SC states, indicating a spin-singlet state.
The discrepancy between the decrease in the spin susceptibility and the absence of the Pauli limit might be explained by the spatially modulated spin susceptibility in the SC2 state.
In addition, NMR measurements revealed that the AFM phase only coexists with the LF SC phase, indicating a strong link to the SC1 state.
Our findings help to understand the unique multi-SC state in CeRh$_2$As$_2$.

\section*{acknowledgments}
The authors would like to thank K. Nogaki, Y. Yanase, Y. Maeno, and S. Yonezawa for their valuable discussions.
This work was partially supported by the Kyoto University LTM Center and Grants-in-Aid for Scientific Research (KAKENHI) (Grants No. JP19K14657, No. JP19H04696, No. JP20H00130, No. JP21K18600, No. JP22H04933, and No. JP23H01124). 
C. G. and E. H. acknowledge support from the DFG program Fermi-NESt through Grant No. GE 602/4-1.
Additionally, E. H. acknowledges funding by the DFG through CRC1143 (project number 247310070) and the Würzburg-Dresden Cluster of Excellence on Complexity and Topology in Quantum Matter—ct.qmat (EXC 2147, project ID 390858490).

S. O. and S. K. contributed equally to this work.

\clearpage
\renewcommand{\figurename}{FIG. S}
\setcounter{figure}{0}

\begin{titlepage}
\begin{center}
\vspace*{12pt}
{\Large Supplementary Materials for Parity transition of \\spin-singlet superconductivity by sub-lattice degrees of freedom}
\vspace{12pt} \\
\end{center}
\end{titlepage}

{\Large Materials and Methods}

\textbf{Sample preparation}
Single crystals of CeRh$_{2}$As$_{2}$ were grown by the Bi flux method.
The present sample is the same as that used in the previous measurements\cite{M.Kibune_PRL_2022SM}.
The sample is in a plate shape with a size of about 3.0 $\times$2.0 $\times$0.75 mm$^3$. 
The superconducting transition temperature $T_{\mathrm{sc}} = 0.37$~K is higher than that in the previous report\cite{S.Khim_Science_2021SM}.

\textbf{AC susceptibility measurements}
The magnetic field dependence of $T_{\rm SC}$ was determined by the onset temperature of the SC diamagnetic signal from the radio-frequency ac susceptibility measurement using an NMR tank circuit.
In the superconducting state, the impedance of the circuit changes due to the superconducting diamagnetic effect, and thus, the tuning frequency of the circuit drastically changes just below $T_{\rm SC}$. 
The frequency of the AC magnetic field is $\sim$ 10~MHz.
Figures~S1 and S2 show the temperature and magnetic field scan of $\chi_{\rm ac}$, respectively.
The obtained phase diagram shown in Fig.~3 (c) is quite consistent with the previous report\cite{S.Khim_Science_2021SM}. 

\textbf{Field alignment}
For NMR measurements, we used a split SC magnet, that generated a horizontal field and combined it with a single-axis rotator to apply a magnetic field parallel to the $c$ axis.
With this setup, we can rotate the magnetic field direction within the $ac$ plane of the sample.

\begin{figure}[!b]
\includegraphics[width=5.5cm,clip]{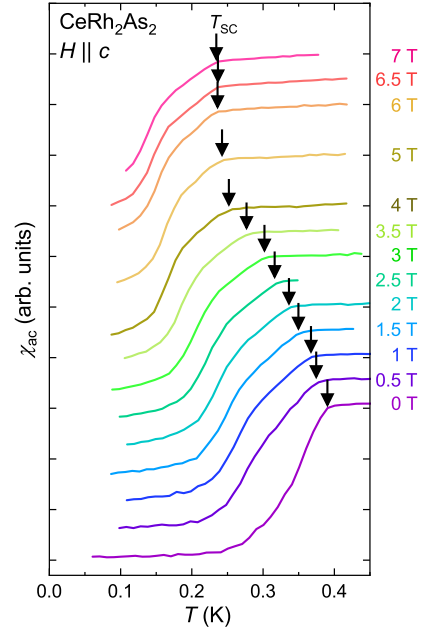}
\caption{
The temperature dependence of $\chi_{\rm ac}$ at several magnetic fields up to 7~T in CeRh$_2$As$_2$.
A magnetic field is applied parallel to the $c$-axis.
The arrows indicate $T_{\rm SC}$ determined by the onset of a superconducting diamagnetic signal.
}
\label{Fig.S1}
\end{figure}

\begin{figure}[!b]
\includegraphics[width=7cm,clip]{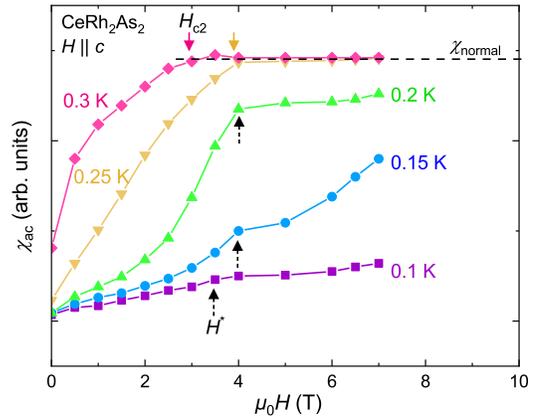}
\caption{
The magnetic field dependence of $\chi_{\rm AC}$ at several temperatures in CeRh$_2$As$_2$.
A magnetic field is applied parallel to the $c$-axis.
The solid arrows indicate $H_{\rm c2}$ and the broken arrows indicate $H^{*}$ where $\chi_{\rm ac}$ shows a kink.
}
\label{Fig.S2}
\end{figure}

\begin{figure*}[!tb]
\includegraphics[width=18cm,clip]{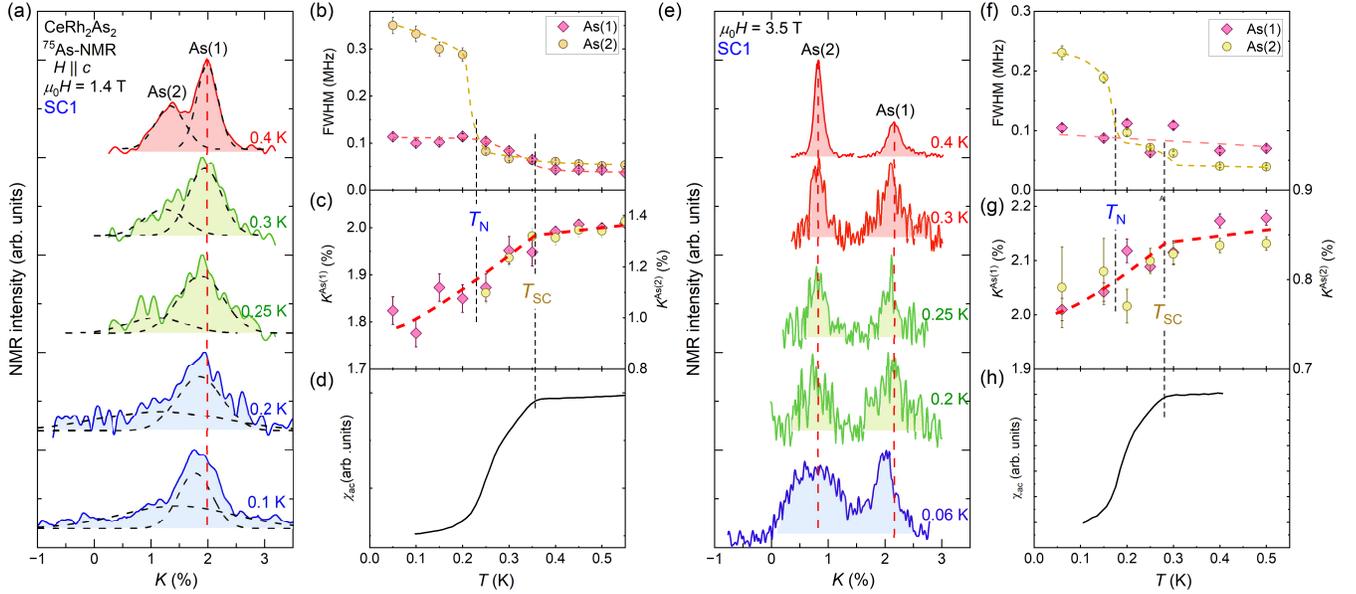}
\caption{
NMR measurements in CeRh$_2$As$_2$.
(a) NMR spectra at 1.4~T for $H \parallel c$ measured at various temperatures.
The dashed curves show Gaussian curves for guidance to the eye.
Temperature dependence of full width at half maximum(FWHM) (b), Knight shift (c) at the As(1) site, $K^{\rm As(1)}$, and As(2) site, $K^{\rm As(2)}$, determined by NMR spectrum at 1.4~T.
$T_{\rm SC}$ and $T_{\rm N}$ are indicated by the dashed lines.
The dashed curves are guides for the eye.
(d) Temperature dependence of ac magnetic susceptibility at 1.4~T for $H \parallel c$.
(e) NMR spectra at 3.5~T for $H \parallel c$ measured at various temperatures.
Temperature dependence of FWHM (f), Knight shift (g), and ac magnetic susceptibility (h) at 3.5~T for $H \parallel c$.
}
\label{Fig.S3}
\end{figure*}

\textbf{NMR measurements}
An NMR spectrometer with a 100 W (at 0 dB input) power amplifier (Thamway Product: N146-5049A) was used for the measurements. 
A conventional spin-echo technique was used for NMR measurements. 
Low-temperature NMR measurements down to 0.06~K were carried out with a $^{3}$He--$^{4}$He dilution refrigerator, in which the sample was immersed into the $^{3}$He--$^{4}$He mixture to avoid radio-frequency heating during measurements. 
We experimentally confirmed the superconductivity just after the NMR pulses using a technique reported in a previous study\cite{K.Ishida_JPSJ_2020SM,H.Fujibayashi_JPSJ_2022SM}.
The $^{75}$As-NMR spectra (nuclear spin $I~=~3/2$, nuclear gyromagnetic ratio $\gamma/2\pi~=~7.29$~MHz/T, and natural abundance 100\%) were obtained as a function of the frequency at fixed magnetic fields.
For $H \parallel c$, because the contribution of the nuclear quadrupole interaction can be ignored at the center peak (1/2 $\leftrightarrow$ -1/2 transition) of the NMR spectrum, the observed NMR spectrum is shown against $K = (f-f_0)/f_0$. 
Here, $f_0 = (\gamma_n/2\pi)\mu_0H$ is the reference frequency.
The magnetic field was calibrated using $^{63}$Cu ($^{63}\gamma_n/2\pi = 11.285$~MHz/T) and $^{65}$Cu ($^{65}\gamma_n/2\pi = 12.089$~MHz/T) NMR signals from the NMR coil. 
The Knight shift is simply determined at the peak of the spectrum.
The anomalies in $T_1$ below $T_{\rm SC}$ were observed in NQR measurements\cite{M.Kibune_PRL_2022SM} and $K$ decreases below $T_{\rm SC}$.
Therefore, we can claim that we are really measuring the superconducting state.

\vspace{5mm}
{\Large Experimental results}

\begin{figure}[!tb]
\includegraphics[width=7.2cm,clip]{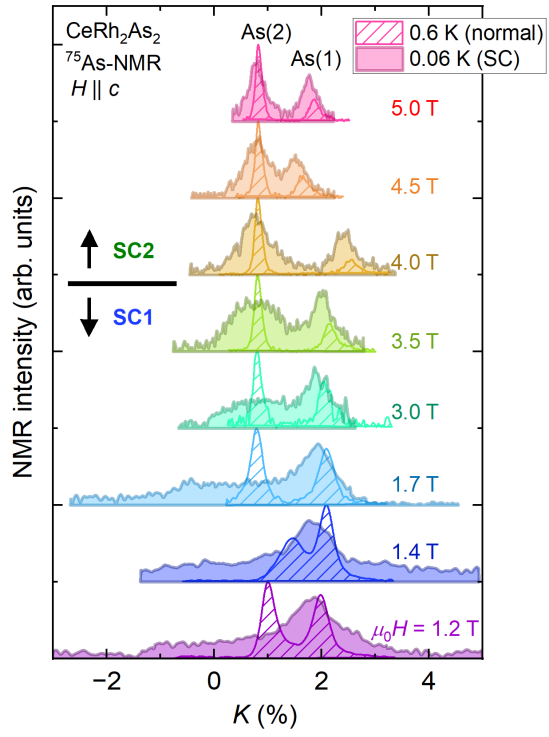}
\caption{
Magnetic field evolution of the NMR spectrum at 0.06~K (superconducting state) and 0.6~K (normal state).
}
\label{Fig.S4}
\end{figure}

\textbf{NMR spectra in the SC1 phase}

To determine the antiferromagnetic transition temperature $T_{\rm N}$ in Fig.~3(c), in addition to 1.2~T and 4.5~T shown in Fig.~2, we also measured the temperature variation of the NMR spectrum at 1.4~T and 3.5~T (both in the SC1 phase) as shown in Fig.~S3.
At 1.4~T and 3.5~T, the clear site-dependent linewidth broadening was observed at low temperatures, indicating the existence of antiferromagnetic transition.
The NMR spectra at 1.4~T overlap even in the normal state, making it difficult to separate the As(1) and As(2) sites at low temperatures.

\textbf{Magnetic field evolution of the NMR spectrum}

Figure S4 shows the magnetic field evolution of the NMR spectrum at 0.6~K (normal state) and 0.06~K (superconducting state).
The magnetic field dependence of the NMR spectrum is affected by a slight misalignment of the magnetic field direction from the $c$-axis ($<$2 degrees
in this experiment).
In situations where the frequency corresponding to the Zeeman energy due
to the external magnetic field (7.29 $\times \mu_0 H$  MHz for As) and
the NQR frequency ($\sim$ 31~MHz and $\sim$ 10.8~MHz for the As(1) and As(2), respectively) are almost the same, a slight
misalignment of the magnetic field direction from the $c$ axis causes the line-width broadening and the change in peak position.
From this figure, we estimated the magnetic field dependence of the full width at half maximum at the lowest temperature and $\Delta K~=$ $K$(0.06~K)-$K$(0.6~K), as shown in Figs.~3(a) and 3(b).
NMR spectra clearly show a decrease in the Knight shift in the superconducting state and the site-dependent linewidth broadening above 4~T (SC2 phase).

\end{document}